\documentstyle[aps,preprint,floats,epsf]{revtex}

\begin{document}

\tighten

\preprint{\vbox{\hbox{JHU-TIPAC-950006}
}}

\title{Theory of Rare $B$ Decays}

\author{Adam F.~Falk}
\address{Department of Physics and Astronomy\\ The Johns Hopkins
University\\ 3400 North Charles Street\\ Baltimore, Maryland 21218
U.S.A.}

\maketitle

\begin{abstract} Theoretical aspects of rare $B$ decays are
reviewed.  The focus is on the relation between short-distance interactions and
physical observables.  It is argued that there remain significant uncertainties
in the theoretical treatment of certain important quantities.
\end{abstract}

\vspace{4cm}

\begin{center}
{Presented at the \\International Symposium on Vector Boson
Self-Interactions,\\
University of California, Los Angeles, California, U.S.A.,\\
February 1--3, 1995.}
\end{center}

\vfill\eject

\section*{Introduction}

While hadrons containing bottom quarks decay weakly, and hence are
quite long-lived, all beautiful things must one day come to an end.  For the
typical $B$ meson, the end comes after about 1.5 ps.  While this provides
enough time for the meson to pass through a measurable
distance within a detector, given today's silicon technology, their
lifetime is still so short that $B$ mesons can be studied
experimentally only by the careful examination of their decay
products.  Hence the study of the bottom quark is essentially the
study of its decays.

\begin{figure}
\epsfxsize=8cm
\centerline{\epsffile{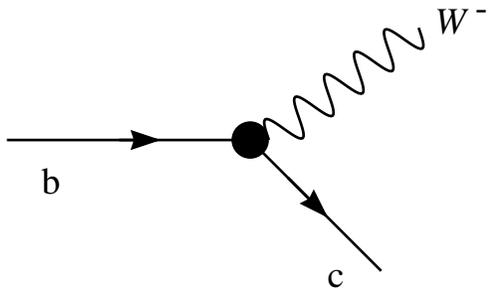}}
\caption{\label{fig2:1}The ordinary decay of a $b$ quark, via $b\to
cW^-$.}
\end{figure}

It is believed that almost all bottom quarks decay weakly into charm
quarks, via the $W$-emission process depicted schematically in
Fig.~\ref{fig2:1}. Rare $b$ decays, then, are those which do not
include the release of a $c$ quark into the final state.  These may
include both Cabibbo-suppressed decays, such as those mediated by
the transition $b\to uW^-$, and flavor-changing neutral decays, such
as penguin-induced transitions.  While the dominant decay mode of
the $b$ quark is believed to be well-understood, it is hoped that
the rare decays may provide a window onto new physics beyond the
standard model.  Not only may one test the standard model by
comparing the small predicted rates for rare channels to experiment,
but the very fact that the charmless channels are suppressed makes
them ideal places to look for anomalous enhancements coming from new
particles and interactions at high energy scales.

\begin{figure}
\epsfxsize=10cm
\centerline{\epsffile{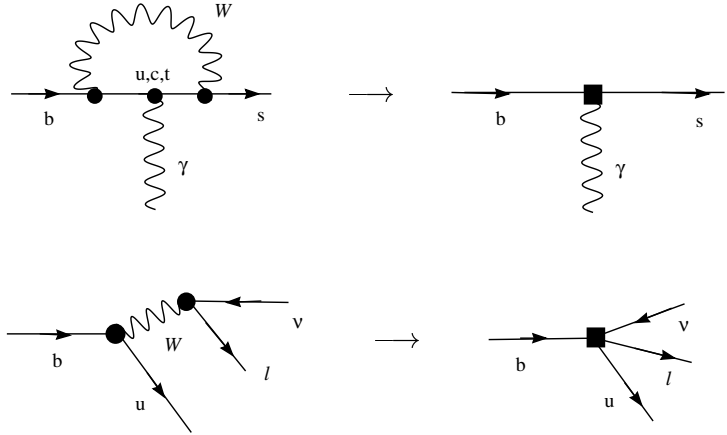}}
\caption{\label{fig2:2-5}Rare decays of a $b$ quark, from the
high-energy and low-energy viewpoints.}
\end{figure}

The theory of rare $b$ decays has two distinct parts, which are
separated from each other conceptually and practically by their
dependence on physics at very different energy scales.  From the
``high-energy'' viewpoint, rare $b$ decays are mediated by
intermediate particles of large virtuality, and the challenge is to
understand the structure of the quark-level transitions which such
virtual particles can induce.  From the ``low-energy'' viewpoint,
rare $b$ decays are mediated by local and nonrenormalizable point
interactions, with coefficients which are determined at high
energies, but at low energies may be viewed simply as coupling
constants of the theory.  The relation between the high-energy and
low-energy viewpoints is demonstrated schematically in
Fig.~\ref{fig2:2-5} for two typical transitions.

{}From the low-energy viewpoint, the theoretical challenge is to
relate the strengths of the suppressed nonrenormalizable quark-level
couplings to {\it physical\/} properties of observable hadrons such
as $B$ and $\Lambda_b$.  The situation is complicated by the
long-distance effects of the strong QCD interactions.  Typically,
the structure of bottom hadrons cannot be computed from first
principles, and one must find techniques which minimize one's
sensitivity to uncomputable low-energy effects, while allowing one
to extract from experiment as much information as possible about
high-energy physics.  At low energies, what one would like to
measure experimentally are the coefficients of the nonrenormalizable
operators such as those pictured in Fig.~\ref{fig2:2-5}.  It is the
goal of low-energy high-energy physics to make this possible.

In what follows, I shall review the theory of rare $b$ decays both
from the high-energy and the low-energy points of view.  In contrast
to the spirit of the rest of this conference, however, my emphasis
will be on the physics at low energies.  In focusing on these
possibly less-familiar effects, I hope to convince this ``high
energy'' audience of the important limitations which low energy
strong interactions place on understanding the physical
manifestations of virtual high energy interactions.  The good news
is that much work is still in progress to minimize these limitations
and to maximize the fundamental discovery potential of experimental
$b$ physics.

\section*{High Energy Viewpoint}

The decays of $b$ quarks, both ordinary and rare, are generated by
virtual interactions at some high scale $M\gg m_b$.  At lower scales
$\mu<M$, these interactions generate nonrenormalizable local
operators.  From the high energy viewpoint, there two questions
which must be answered:

1. What operators are generated?

2. With what coefficients?
\begin{figure}
\epsfxsize=10cm
\centerline{\epsffile{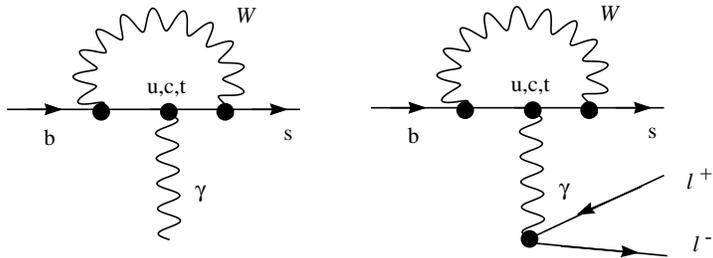}}
\caption{\label{fig3:1-2}Penguin-induced decays of the $b$ quark.}
\end{figure}
\par\noindent For example, the penguin diagrams pictured in
Fig.~\ref{fig3:1-2} generate, among others, the operators
\begin{eqnarray}
   C_7\,{\cal O}_7&=&C_7(\mu)\,(\bar s\sigma^{\mu\nu}b)_R\,
F_{\mu\nu}\,,\nonumber\\
   C_8\,{\cal O}_8&=&C_8(\mu)\,(\bar s\gamma^\mu
b)_L\,\bar\ell\gamma_\mu\ell\,,\\
   C_9\,{\cal O}_9&=&C_9(\mu)\,(\bar s\gamma^\mu b)_L\,
   \bar\ell\gamma_\mu\gamma_5\ell\,.\nonumber
\end{eqnarray} Perturbative QCD corrections are included by dressing
the graphs in Fig.~\ref{fig3:1-2} with gluons.  The leading
logarithm approximation, which resums all terms of the form
$\alpha_s^n(\mu)\ln^n(\mu/M_W)$, suffers from a strong ambiguity in the
choice of renormalization scale $\mu$.  The resolution of this ambiguity will
only properly be resolved by a full next-to-leading order calculation.  The
present state
of the art for the coefficient $C_7(\mu)$ is summarized in
Ref.~\cite{CFMRS}.  The calculation is complete to order $\alpha_s$,
and partially complete at next-to-leading order.  Varying the
renormalization scale $\mu$ from $m_b/2$ to $m_b$, one finds a
residual scale-dependence uncertainty of approximately $\pm15\%$.
Since ${\cal O}_7$ is the operator which is primarily responsible
for the rare decay $B\to X_s\gamma$, there is a corresponding
uncertainty of at least $\pm30\%$ in the prediction of this decay rate in the
standard model.  The coefficients $C_8(\mu)$ and
$C_9(\mu)$, which are responsible for the decay $B\to
X_s\ell^+\ell^-$, are known with similar accuracy.

\begin{figure}
\epsfxsize=8cm
\centerline{\epsffile{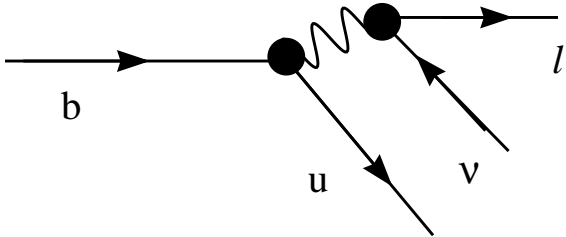}}
\caption{\label{fig4:1}Charmless semileptonic $b$ quark decay.}
\end{figure}

There are also charmless weak $b$ decays, which are rare because
their rates are suppressed compared to the dominant weak decay mode by
the factor
$|V_{ub}/V_{cb}|^2\sim10^{-2}$.  Charmless semileptonic decays,
shown in Fig.~\ref{fig4:1}, arise from operators of the form
\begin{equation}
   A(\mu)\bar
u\gamma^\mu(1-\gamma_5)b\,\bar\ell\gamma_\mu(1-\gamma_5)\nu\,.
\end{equation}
 Since this operator may be written, up to weak and
electromagnetic corrections and fermion masses, as a product of
conserved currents, the coefficient $A(\mu)$ suffers from no scale
ambiguity.  It has been computed to order $\alpha_s(m_b)$, and the
residual uncertainty is small.

\begin{figure}
\epsfxsize=8cm
\centerline{\epsffile{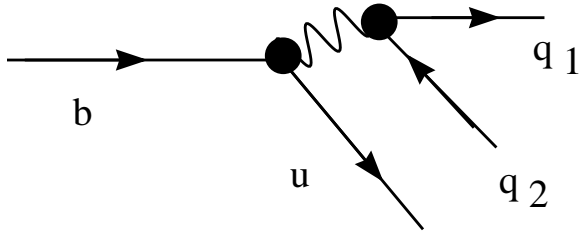}}
\caption{\label{fig4:2}Charmless nonleptonic $b$ quark decay.}
\end{figure}

The same is not true of charmless nonleptonic decays, mediated by
operators such as shown in Fig.~\ref{fig4:2}.
 At low energies, these diagrams induce four-quark
operators of the form
\begin{eqnarray}
   C_1\,{\cal O}_1&=&C_1(\mu)\,\bar u_i\gamma^\mu(1-\gamma_5)b_i\,
   \bar q_{1j}\gamma_\mu(1-\gamma_5)q_{2j}\,,\nonumber\\
   C_2\,{\cal O}_2&=&C_2(\mu)\,\bar u_i\gamma^\mu(1-\gamma_5)b_j\,
   \bar q_{1j}\gamma_\mu(1-\gamma_5)q_{2i}\,,\\
\end{eqnarray} where the indices $i$ and $j$ indicate sums over
colors.  These operators are {\it not\/} products of currents, and
they receive renormalizations from perturbative QCD which are as
large as those received by penguin operators.  As summarized in
Ref.~\cite{Buras}, the coefficients
$C_1(\mu)$ and $C_2(\mu)$ have now been computed at next-to-leading
order.  The residual uncertainty, largely arising from
scheme-dependence, is about $\pm15\%$.

\begin{figure}
\epsfxsize=14cm
\centerline{\epsffile{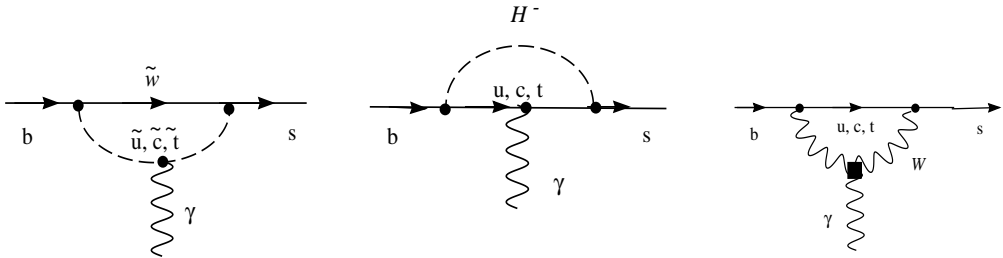}}
\caption{\label{fig5:1-3}New physics contributions to the rare
process $b\to X_s\gamma$.}
\end{figure}

New physics at high energies can also contribute to the coefficients
$C_i(\mu)$.  For example, as shown in Fig.~\ref{fig5:1-3},
supersymmetric particles, extra scalars, and anomalous trilinear
gauge couplings can all modify
$C_7(\mu)$ as compared to the standard model.
  The size of these new contributions depends, of course,
on the particular model involved.  However, because of the
uncertainties inherent in the perturbative corrections to the
standard model, new physics will {\it only\/} be observable in rare
$b$ decays if causes deviations from the standard model at
significantly more than the 15\% level.  This is the most important
lesson to be taken from the high energy point of view.

\section*{Low Energy Viewpoint}

At low energies, we start with an interaction Lagrangian density
which is a sum over nonrenormalizable operators with coefficients
determined at high energies,
\begin{equation}\label{lagrangian}
   {\cal L} = \sum C_i(\mu)\,{\cal O}_i(\mu)\,.
\end{equation} The matrix elements of the operators ${\cal O}_i$ are
defined so as to cancel the $\mu$ dependence of any physical
observable.  The operators and their coefficients are renormalized
at a low energy scale $\mu\sim m_b$, and nonrenormalizable terms are
suppressed by powers of the scale $M$ at which the interactions
become nonlocal and new physics comes into play.

The challenge, at low energies, is to use the Lagrangian
(\ref{lagrangian}) to make {\it physical\/} predictions.  One option
is to try to predict {\it exclusive\/} decay modes, such as $B\to
K^*\gamma$ or $B\to\rho\ell\nu$.  However, the theoretical methods
available are not entirely satisfactory:  the Heavy Quark Effective
Theory, so useful for $b\to c$ transitions~\cite{HQET}, is of
limited applicability here with only light quarks in the final
state.  Lattice calculations eventually may provide important
information on exclusive matrix elements, but that is for the most
part still in the future.  For now, one is left to rely on phenomenological
models, which for all their occasional successes do not provide any controlled
approximation to QCD.

Alternatively, one may consider {\it inclusive\/} decay modes, such
as $B\to X_s\gamma$ or $B\to X_u\ell\nu$.  There has been
considerable recent progress~\cite{inclusive} in the computation of
such quantities in a simultaneous expansion in $1/m_b^n$ and
$\alpha_s(m_b)^n$.  It has recently been understood, as well, that
there are important limitations to such calculations.  We shall now review this
situation in some detail.

The theoretical analysis of inclusive $B$ decays relies on the
Operator Product Expansion and perturbative QCD.  The partial width
$\Gamma$ for an operator ${\cal O}$ to mediate the decay of a $B$ to any final
state $X$ with the correct quantum numbers is proportional to the square of the
matrix element, summed over the possible final states,
\begin{equation}
   \Gamma\sim\sum_X\big|\langle X|\,{\cal O}\,|B\rangle\big|^2\,.
\end{equation}
By the Optical Theorem, $\Gamma$ may be rewritten as
the imaginary part of a forward scattering amplitude,
\begin{equation}\label{top}
   \Gamma\sim{\rm Im}\,\langle B|\,T\{{\cal O},{\cal
O}^\dagger\}\,|B\rangle\,,
\end{equation}
which is then expanded simultaneously in powers of
$\alpha_s(m_b)$ and $1/m_b$.  One obtains expressions for the inclusive partial
widths; for example~\cite{inclusive,charmless},
\begin{eqnarray}\label{partialwidths}
   \Gamma(B\to
   X_s\gamma)&\propto&m_b^5\left|C_7(\mu)\right|^2\bigg\{
   1+{\lambda_1-9\lambda_2\over2m_b^2}+K(\mu)\,\alpha_s(m_b)
   +\ldots\bigg\}\,,\nonumber\\\Gamma(B\to
   X_u\ell\nu)&\propto&m_b^5\left|V_{ub}\right|^2\bigg\{
   1+{\lambda_1+3\lambda_2\over2m_b^2}
   +K_{\rm s.l.}\,\alpha_s(m_b)+\ldots\bigg\}\,.
\end{eqnarray} Here the nonperturbative parameters $\lambda_1$ and
$\lambda_2$ are defined by hadronic matrix elements~\cite{FN},
\begin{eqnarray}
   \lambda_1&=&\langle B|\,\bar b(iD)^2b\,|B\rangle/2m_b\,,
   \nonumber\\
   \lambda_2&=&\langle B|\,\bar b(-{\textstyle{i\over2}}
   \sigma^{\mu\nu})G_{\mu\nu}b\,|B\rangle/2m_b\,.
\end{eqnarray}
It is straightforward to find $K_{\rm
s.l.}={2\over3\pi}({25\over4}-\pi^2)$~\cite{GPR}, while the perturbative
correction $K(\mu)$ is too messy to be
illuminating~\cite{CFMRS}.

There are a number of sources of uncertainty in the
expressions~(\ref{partialwidths}).  The operator $\bar
b(-{\textstyle{i\over2}}\sigma^{\mu\nu})G_{\mu\nu}b$ violates the
Heavy Quark Spin Symmetry and may be measured directly from the
$B$--$B^*$ mass difference,
$\lambda_2\approx0.12\,{\rm GeV}^2$.  However, $\lambda_1$ can not
be measured directly; instead, one must rely on phenomenological
models.  While this is unfortunate, if we assume $\lambda_1\le1\,{\rm
GeV}^2$, then inspection of Eq.~(\ref{partialwidths}) shows that the
error induced in the partial widths $\Gamma$ is $10\%$ or less.  Higher order
nonperturbative corrections, of order $1/m_b^3$, are expected to be at the
level of a few percent.

The primary sources of uncertainty in Eq.~(\ref{partialwidths}) are
the value to take for the bottom mass $m_b$, and uncomputed higher
order radiative corrections.  Because of the overall factor of
$m_b^5$, the theoretical partial widths are extremely sensitive to
this parameter.  For example, allowing $m_b$ to vary over the range
$4.5\,{\rm GeV}\le m_b\le5.0\,{\rm GeV}$ induces an uncertainty in
$\Gamma$ of approximately $50\%$.  While lower values for $m_b$ seem
currently to be preferred, the issue is still quite unsettled.\footnote{There
is an ongoing controversy over issues as fundamental as the proper {\it
definition\/} of $m_b$.  I will not review this discussion here.}

\subsection{Higher order radiative corrections}

\begin{figure}
\epsfxsize=10cm
\centerline{\epsffile{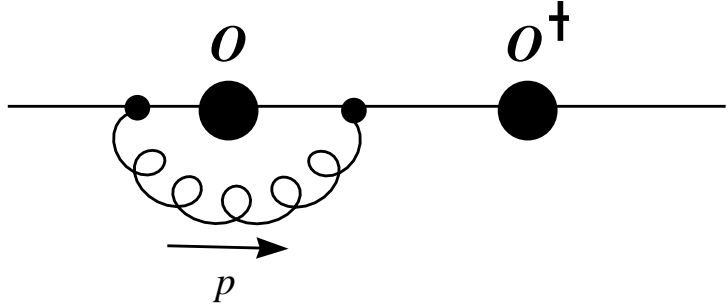}}
\caption{\label{fig8:1}A typical one-loop radiative correction to
$T\{{\cal O},{\cal O}^\dagger\}$.}
\end{figure}

\begin{figure}
\epsfxsize=13cm
\centerline{\epsffile{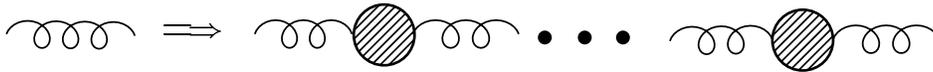}}
\caption{\label{fig8:2}The gluon propagator replaced by the sum of
self-energy graphs.}
\end{figure}

Higher order radiative corrections to the partial widths have
recently been considered by a number of authors.  In particular,
attention has been paid to a set of corrections which are dominant
in the limit of large $N_f$ (large number of quark flavours), and
which in the real world still may be particularly large.  These come from
taking the one loop radiative correction to the time-ordered
product~(\ref{top}), an example of which is shown in
Fig.~\ref{fig8:1}, and replacing the gluon propagator with a sum of
self-energy bubbles.
 If this replacement, which is illustrated in
Fig.~\ref{fig8:2}, is carried out to all orders and then extrapolated
to the physical $N_f$, it amounts to replacing the strong coupling
constant $\alpha_s(m_b)$ by its running value
$\alpha_s(p^2)$ evaluated at the loop momentum.

The BLM scale-setting prescription~\cite{BLM} requires that one
perform the substitution shown in Fig.~\ref{fig8:2} to leading order;
the two-loop contribution to the radiative correction is then
expected on general grounds to be parametrically large.  Once this
part of the two-loop computation has been done, one adjusts the
scale $\mu$ in the one-loop result to absorb it.  A recent
application of this criterion to the inclusive rate for $B\to
X_u\ell\nu$ indicates that the appropriate scale for this process is
$\mu\sim m_b/10$ rather than $m_b$~\cite{LSW}.  A more complicated
scale-setting procedure which resums all orders in the bubble sum
(but which suffers from a certain lack of uniqueness) does not, in
general, lead to quite such a low value of $\mu$~\cite{Neubert},
although the two-loop corrections are, of course, still quite large.

This treatment of higher order radiative corrections leaves us with
two questions.
\par\noindent 1. Should such a low renormalization scale be taken
seriously?  If so, then clearly the entire program of computing
inclusive rates perturbatively is in trouble.  If not, then one
still has to do deal with the fact that two-loop corrections are
much larger than one might na\"\i vely have thought.
\par\noindent  2. If there is a class of diagrams which is unusually
large, can perturbation theory be improved in a sensible way?  Once
such a resummation has been performed, can one show that the
remaining uncertainties are likely to be small?

\subsection{The need for endpoint spectra}

\begin{figure}
\epsfxsize=13cm
\centerline{\epsffile{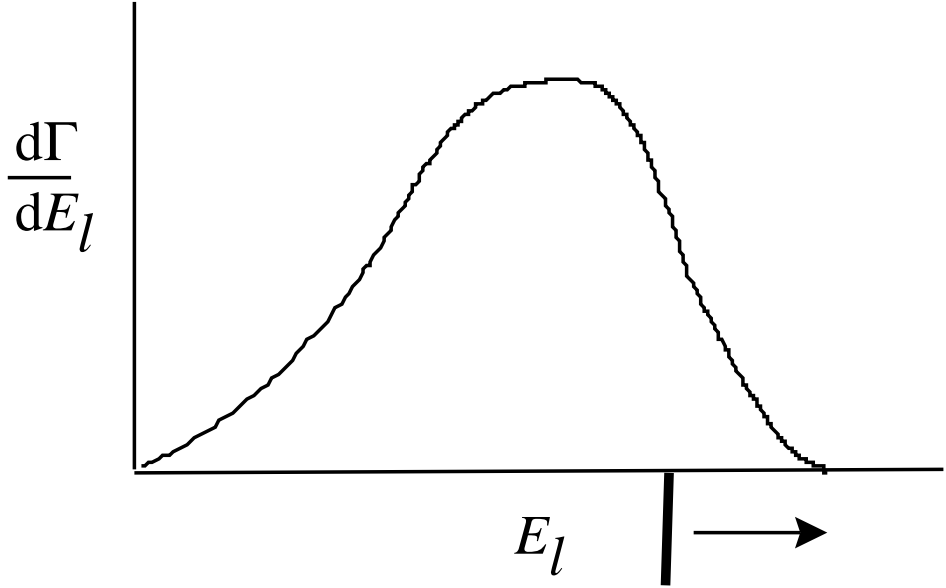}}
\caption{\label{fig9:1}A cartoon of the lepton energy spectrum in
$B\to X_u\ell\nu$, along with the kinematic cut.}
\end{figure}

Final states with charm present an enormous background to rare $B$
decays.  For example, the decay $B\to X_c\ell\nu$ obscures $B\to
X_u\ell\nu$, and $B\to D\pi^0\to D\gamma\gamma$ presents a
problematic background to $B\to X_s\gamma$.  Typically, strict
kinematic cuts are used to exclude such process.  For example,
studies of rare decays accept only leptons and photons with energies in the
range $2.2\,{\rm GeV}\le E_\ell, E_\gamma\le2.7\,{\rm GeV}$, beyond the
kinematic endpoint for charm in the final state.  Hence it is necessary for
theorists to compute not only partial widths $\Gamma$, but inclusive lepton
{\it spectra\/} ${\rm
d}\Gamma/{\rm d}E$ within $20\%$ or so of the endpoint.  A cartoon
of a lepton energy spectrum, along with the kinematic cut, is shown
in Fig.~\ref{fig9:1}.

The theoretical problem is that the OPE does not converge when
resticted to the lepton or photon energy endpoint.  What is
computable is not the full differential spectrum ${\rm d}\Gamma/{\rm
d}E$ but rather {\it moments\/} of this spectrum, obtained by
weighting the differential spectrum by some function and then
integrating.  Introducing the scaled energy variable
$y=2E_{\ell,\gamma}/m_b$, we thus ``smear'' with a weighting
function with support only in the small region $1-\delta\le
y\le1$.  The size $\delta$ of the smearing region controls the
convergence of the OPE.  For $\delta\sim\Lambda_{\rm
QCD}/m_b\sim10\%$, {\it all\/} orders in the $1/m_b$ expansion
contribute equally~\cite{shape}, and the leading terms in the
expansion~(\ref{partialwidths}) is clearly insufficient.  Of
course, this is only an order of magnitude estimate, and how the OPE
converges for the experimentally chosen upper value of $\delta$
cannot be determined from such general considerations.  At this
point, then, a certain amount of faith is required in the
interpretation of the smeared theoretical spectra.

An interesting by-product of this analysis is the result that the
same infinite sum of terms in the $1/m_b^n$ expansion determines
the shape of the endpoint spectrum in $B\to X_s\gamma$ and in $B\to
X_u\ell\nu$~\cite{shape,FJMW}.  Whether this relation yields
useful predictive power is still to be seen.

\subsection{Sudakov Logarithms}

Another source of uncertainty in the shape of the endpoint spectrum comes from
Sudakov logarithms~\cite{Sudakov}.  For example, the perturbative corrections
to the lepton energy spectrum in $B\to X_u\ell\nu$ is extremely singular near
the endpoint $y=1$~\cite{JK}:
\begin{equation}
   {{\rm d}\Gamma\over{\rm d}y}={{\rm d}\Gamma_0\over{\rm d}y}
   \bigg\{ 1-{2\alpha_s\over3\pi}\bigg[\ln^2(1-y)+{31\over6}\ln(1-y)
   +\ldots\bigg]+{\cal O}(\alpha_s^2)\bigg\}\,,
\end{equation}
where ${\rm d}\Gamma_0/{\rm d}y$ is the spectrum at tree level.  At order
$\alpha_s^2$, the leading singularity is $\ln^4(1-y)$, and so forth.  These
Sudakov double logarithms may be resummed into an exponential suppression
factor:
\begin{equation}
   {{\rm d}\Gamma\over{\rm d}y}={{\rm d}\Gamma_0\over{\rm d}y}
   \exp\bigg\{-{2\alpha_s\over3\pi}\ln^2(1-y)\bigg\}+\ldots\,.
\end{equation}
This leading behaviour is actually {\it stronger\/} very near $y=1$ than that
given by the nonperturbative power corrections, but it is calculable.

What must be suppressed are the leading {\it uncalculated\/} corrections, which
is accomplished by smearing over a region $\delta$ large enough that they may
be neglected.  In the large $m_b$ limit, this requires that we smear over a
region formally much larger than $\delta\sim\Lambda_{\rm QCD}/m_b$, given by
the condition~\cite{FJMW}
\begin{equation}
   \delta>\exp\bigg\{-\sqrt{\pi/\alpha_s(m_b)}\bigg\}\,.
\end{equation}
In this strict limit, then, all nonperturbative corrections to the endpoint
shape would be irrelevant.  But for realistic $m_b\approx4.8\,{\rm GeV}$,
and the given experimental smearing region $\delta\approx0.1\sim0.2$, do the
uncalculated Sudakov effects {\it actually\/} dominate the nonperturbative
power corrections?  It is difficult to guess, based only on na\"\i ve power
counting arguments.  Explicit calculations of the subleading Sudakov logarithms
may help clarify the situation.

For technical reasons, the Sudakov corrections to the smeared photon spectrum
in $B\to X_s\gamma$ are under much better theoretical control than in $B\to
X_u\ell\nu$~\cite{shape}.  They do not introduce unmanageable uncertainties
into the computation of the weighted spectra.

\subsection{Instantons}

\begin{figure}
\epsfxsize=10cm
\centerline{\epsffile{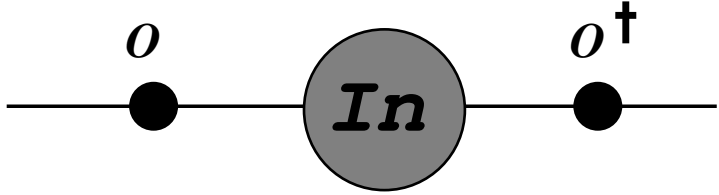}}
\caption{\label{fig11:1}The one instanton contribution to $T\{{\cal O},{\cal
O}^\dagger\}$.}
\end{figure}

Finally, there are possibly large contributions to energy endpoint spectra from
instantons.  These arise because the light quark which is produced in the
short-distance interactions can propagate in an instanton background, as
pictured in Fig.~\ref{fig11:1}.
Chay and Rey computed, in the dilute instanton gas approximation, the one
instanton contribution to ${\rm d}\Gamma/{\rm d}y$ for $B\to X_u\ell\nu$ and
$B\to X_s\gamma$~\cite{CR}.  Their result diverges dramatically at the
endpoint, as $y\to1$.  The contribution to $B\to X_s\gamma$ is nonetheless
small and under control when one computes weighted spectra, but the same is not
true for $B\to X_u\ell\nu$.  Instead, one finds that the one instanton
contribution is entirely untrustworthy in the experimentally defined window.

The one instanton contribution goes bad in this region presumably because
multi-instanton configurations begin to be important.  We have used the one
instanton calculation as the motivation for a {\it crude\/} ansatz for the
multi-instanton result in this region~\cite{FK}.  This ansatz incorporates, as
much as possible, the reliable information from the one-instanton calculation.
When we vary this na\"\i ve ``best guess'' ansatz by two orders of magnitude,
we find that over most of the ansatz parameter space, the instantons do in fact
dominate the weighted endpoint spectra.

One must be careful about interpreting this result.  It is potentially
interesting only in a {\it negative\/} sense.  On the one hand, the actual
numbers certainly cannot be believed; by no means do we claim to have computed
the correct multi-instanton contribution.  On the other, we have failed to find
any justification for ignoring the instantons in the endpoint region.  In light
of this equivocal situation, one may well wonder whether one can still trust
the relationship between the endpoint spectra for $B\to X_s\gamma$ and $B\to
X_u\ell\nu$ proposed in Refs.~\cite{shape,FJMW}.  This is a situation badly in
need of clarification.  Invocations of faith, one way or the other, will not be
sufficient; a more sophisticated estimate of instanton contributions is what is
required.  Such an estimate could show, for example, that our ansatz for the
multi-instanton contribution is entirely too crude, and that other techniques
can be used to prove that the multi-instanton contribution is necessarily
negligible.  We certainly hope that this will prove to be the case.

\section*{Conclusions}

We may summarize the status of the theory of rare $B$ decays from each of our
two viewpoints:
\smallskip
\par\noindent {\bf Low Energy Viewpoint:}
\smallskip
\par\noindent 1. Exclusive decay rates are extremely difficult to compute
reliably.  One must resort to models and other uncontrolled assumptions, a
situation which is most unsatisfactory.
\par\noindent 2. Inclusive calculations, by contrast, may be performed in a
controlled expansion in powers of $\alpha_s$ and $1/m_b$.  However, there
remain unresolved uncertainties about
\par a. uncomputed higher orders in $\alpha_s$ and the renormalization scale
$\mu$;
\par b. Sudakov double logarithms near the lepton energy endpoint;
\par c. instanton contributions near the lepton energy endpoint.
\par\noindent 3. The decay $B\to X_s\gamma$ is in much better shape with
respect to Sudakov and instanton corrections than is $B\to X_u\ell\nu$.  Hence,
while the calculation of $B\to X_s\gamma$ is itself perhaps fairly secure, the
proposed relationship between the endpoint spectra in $B\to X_s\gamma$ and
$B\to X_u\ell\nu$ may well be threatened by these effects.
\smallskip
\par\noindent {\bf High Energy Viewpoint:}
\smallskip
\par\noindent 1. In view of the significant uncertainties in existing
theoretical calculations, only modifications to the Standard Model which affect
rare decays at the $50\%$ level or higher are likely to be experimentally
detectable.  Small modifications, say at the $10\%$ level, are unlikely ever to
be seen.
\par\noindent 2. There is considerable room for the situation to improve, and
much work remains to be done.

\section*{Acknowledgments}

It is a pleasure to thank the organizers for a stimulating
conference and absolutely lovely weather.  This work was supported
by the National Science Foundation under Grant No.~PHY-9404057 and
National Young Investigator Award No.~PHY-9457916, and by the
Department of Energy under Outstanding Junior Investigator Award
No.~DE-FG02-094ER40869.

\end{document}